\documentclass[fleqn,10pt]{wlscirep}
\usepackage[utf8]{inputenc}
\usepackage[T1]{fontenc}
\usepackage{physics}
\usepackage[utf8]{inputenc}
\usepackage[T1]{fontenc}
\usepackage{bm}
\usepackage{cite}
\usepackage{amsmath,amssymb,amsfonts}
\usepackage{algorithmic}
\usepackage{graphicx}
\usepackage{textcomp}
\usepackage{xcolor,stfloats}
\usepackage{caption}
\usepackage{subcaption}
\usepackage{setspace}
\usepackage{physics}
\usepackage{lineno}

\title{Monolithic Integration of Quantum Resonant Tunneling Gate on a 22nm FD-SOI CMOS Process} 

\author[1,*,+]{Imran Bashir}
\author[1,+]{Dirk Leipold}
\author[2,3,*,+]{Elena Blokhina}
\author[1,+]{Mike Asker}
\author[2]{David Redmond}
\author[2]{Ali Esmailiyan}
\author[2]{Panagiotis Giounanlis}
\author[1,+]{Hans Haenlein}
\author[2]{Xutong Wu}
\author[2,3]{Andrii Sokolov}
\author[3] {Andrew K. Mitchell}
\author[2,3]{Dennis Andrade-Miceli}
\author[1,2]{Robert Bogdan Staszewski}
\affil[1]{Equal1.labs, Freemont, California, USA}
\affil[2]{Equal1.labs, NovaUCD, Belfield, Dublin, Ireland}
\affil[3]{University College Dublin, Belfield, Dublin, Ireland}

\affil[*]{Corresponding Authors: imran.bashir@\{equal1.com, ieee.org\}, elena.blokhina@\{equal1.com, ucd.ie\}, }

\begin{abstract}

The proliferation of quantum computing technologies has fueled the race to build a practical quantum computer. The spectrum of the innovation is wide and encompasses many aspects of this technology, such as the qubit, control and detection mechanism, cryogenic electronics, and system integration. A few of those emerging technologies are poised for successful monolithic integration of cryogenic electronics with the quantum structure where the qubits reside. In this work, we present a fully integrated Quantum Processor Unit in which the quantum core is co-located with control and detection circuits on the same die in a commercial 22-nm FD-SOI process from GlobalFoundries. The system described in this work comprises a two dimensional (2D) 240 qubits array integrated with 8 detectors and 32 injectors operating at 3\,K and inside a two-stage Gifford\,-McMahon cryo-cooler. The power consumption of each detector and injector is 1\,mW and 0.27\,mW, respectively. The control sequence is programmed into an on-chip pattern generator that acts as a command and control block for all hardware in the Quantum Processor Unit. Using the aforementioned apparatus, we performed a quantum resonant tunneling experiment on two qubits inside the 2D qubit array. With supporting lab measurements, we demonstrate the feasibility of the proposed architecture in scaling-up the existing quantum core to thousands of qubits.

\end{abstract}
\begin{document}

\flushbottom
\maketitle

\section{Introduction}

Several quantum computing technologies are making inroads in the market today. The most commercially prevalent is the superconducting transmon qubit where a flux tunable critical current in a Josephson junction loop~\cite{knight_2017,JKelley_2017} or superconducting quantum interference device (SQUID) controls the Pauli rotation about an axis on the Bloch sphere. The apparatus driving the transmon qubit is expansive (Fig. 1 in ref.~\cite{bogdan2021}). In order to reduce the thermal excitation of the system, the transmon is placed in a dilution fridge operating at a base temperature of 10\,mK. In this state, the flux-tunable transmon behaves like a quantum mechanical system where the driving signal amplitude or phase controls the XY rotation.
The drive signal is generated by a rack mount arbitrary waveform generator (AWG) which is passed through mixers, attenuators, and couplers. The transmon qubit also includes a DC current port to tune the qubit resonance frequency and the current into that port is sourced by bias DACs. Lastly, the transmon qubit has a read-out port which is AC-coupled to a resonator. The qubit's state is determined by coupling an RF signal to the resonator and measuring the amplitude or phase shift of the reflected signal. The RF signal is coupled to the qubit through circulators and attenuators, while the reflected signal is amplified in multiple stages before being sourced to a measurement system. All the rack-mount control equipment such as AWGs, bias DACs, signal generators, and signal analyzers are located at room temperature whereas the couplers, mixers, attenuators, and circulators are strategically placed at multiple stages in the dilution fridge. Specialized coax cables for RF signals are used at each stage to minimize active and passive heating, especially at the base temperature stage where the thermal load of the dilution fridge (such as Bluefors XLD400 DR) is 19$\mu$W~\cite{krinner18}. This apparatus works reasonably well for small qubit counts (50-100) where the reported error rate is on the order of 0.1$\%$~\cite{barends18} and the qubit relaxation time ($\rm T_1$) is 18.3$\mu$s~\cite{bardin_2019b}. However, a fault-tolerant system requires a significantly large number of qubits which poses a challenge to the aforementioned approach when considering the amount of equipment, cabling, active and passive components required to facilitate control and read-out operation on such a quantum processor. Other challenges to scaling this technology to millions of qubits include frequency collision and crowding~\cite{Brink_2018}.

\begin{figure}[t!]
     \centering
        \includegraphics[width=\textwidth]{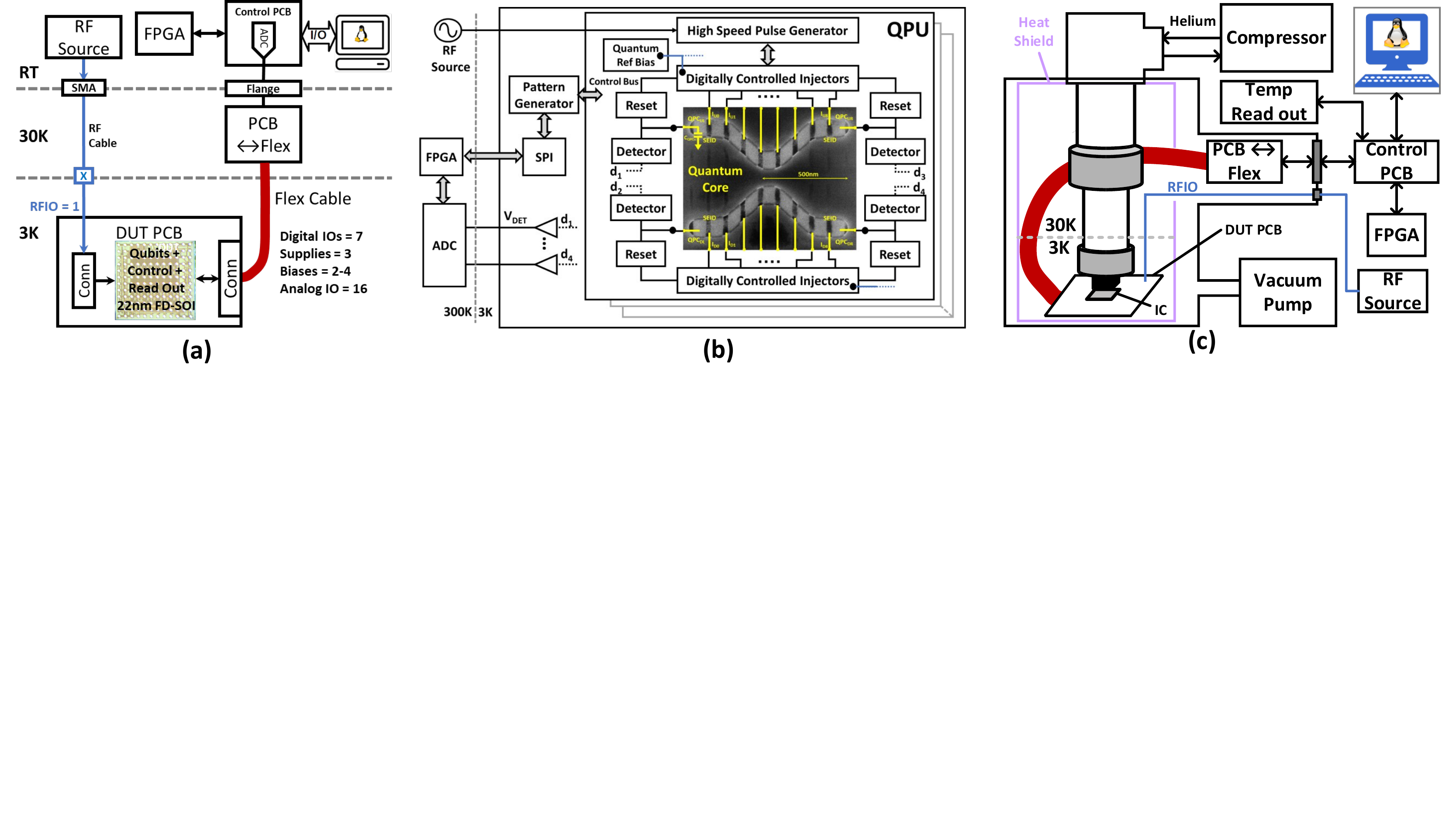}
        \vspace{-60mm}
        \caption{{\bf The quantum computer apparatus and top-level architecture.} (a) Complete functional view of the quantum computer with cabling. The breakdown of the wiring on the flex cable is also shown. The 16 analog I/O wires are routed to the eight (8) on-chip detectors and can be eliminated with on-chip A/D. (b) Block level architecture of the Quantum Processor Unit. The 2D quantum structure, shown as (transmission electron microscopy) TEM micrograph, is co-located with control and read-out electronics on the same die. (c) Graphical representation of the Cryocooler apparatus and the interface between the IC and room temperature electronics.}
        \label{fig:cryosys}
\end{figure}

On the other hand, semiconductor qubits offer a more efficient path to scaling \cite{nikandish21}. In this technology, an electron's position or spin is manipulated while they are confined in nanoscopic-scale quantum dots~\cite{Maurand_2016}. The number of qubits is scaled by placing a large number of quantum dots in an array. Pauli rotations are typically performed by driving a microwave signal and/or appropriate DC biasing to a control node to manipulate the electron's tunneling or spin blockade process while the qubits are placed in a dilution fridge with a base temperature of 20\,mK. Read out can be done by sensing charge, current, or phase (for a gate-dispersive read out~\cite{crippa_nature19}). The key benefit of this approach is the advanced lithography process with which quantum dots with diameters as low as 45\,nm~\cite{zwerver21} have been fabricated. However, the decoherence process of such qubits can be influenced by the nanoscopic charge fluctuation from the $^{28}$Si/$^{28}$SiO$_2$ interface. That said, the qubits manufactured using this process have achieved a spin relaxation time ($\rm T_1$) of 1.6\,s, a spin dephasing time ($\rm T_2$) of 24\,$\mu$s and single qubit gate fidelity of 99\,$\%$~\cite{zwerver21}. Scaling the spin qubits count to millions will pose a challenge to route a large number of drive channels to the qubit array, especially when the control equipment resides at room temperature. The same will be true for detection channels if a gate-dispersive read-out method is used.

The obvious solution to cable management of a large number of wide-bandwidth I/O channels for qubit operation is to simplify the interface by deploying the control and detection hardware inside the cryocooler and closer to the qubit substrate. This complex integration task requires intimate knowledge of the cooling power and active and passive heat dissipation from circuits and cabling at each stage in the cryocooler. The first stage of the dilution fridge typically has a pulse tube cryocooler operating at 3\,K with a thermal load specification of 1.5\,W. This would be an ideal environment for a controller chip capable of driving a large number of excitation signals into a qubit processor while the number of wires to and from the room temperature electronics is reduced to a few control and trigger lines. Such controllers have been reported for transmon qubits~\cite{bardin_2019b} and spin qubits~\cite{dijk_2020,Bonen_2018}. The same concept has been applied to cryogenic detectors for spin qubits using gate-dispersive read-out~\cite{ruffino21} and transimpedance amplifiers\cite{Voinigescu_2019}. In another published integration scenario, the qubit substrate and the controller IC is bonded to a common substrate at an elevated temperature stage of 100\,mK~\cite{pauka21} while the driving circuits generating the excitation signal used a switch capacitor topology. The typical cooling power at this stage is 200\,$\mu$W, which is ten times higher than the 20\,mK stage, however it is not large enough to accommodate meaningful integration with large number of channels. In another published work, the controller chip with single channel driver and detector was integrated with a double quantum dot structure on the same die~\cite{Guevel_2020} operating at 110\,mK. That solution cannot be scaled to a large qubit count due to the limited cooling power at the 110\,mK stage.


This concludes the discussion on prior art and their unique characteristics. The benefits of the approach suggested in this paper and their distinction with the aforementioned references will be discussed in detail in the next section.

\section{Key Innovation Factors}
\label{sec:qc_app1}

The architecture that will be described in this paper is unique from the prior art in several ways. Firstly, the cryogenic quantum control and read-out system is integrated with a two-dimensional (2D) qubit array in a single chip operating at 3\,K as shown in Fig.~\ref{fig:cryosys}(a). The result is a simplified and seamless interface between the cryogenic electronics and the qubit array. Only 30 wires are routed over a ``flex" cable to the apparatus at room temperature. In addition to the flex cable, a single RF coax cable connects the IC to an external signal generator to receive the system clock. The digitization of the read-out signal is currently off-chip but can be straightforwardly relocated on-chip, reducing the wire count to the room temperature apparatus to just 14. The thermal load of entire apparatus shown in Fig.~\ref{fig:cryosys}(a) is 21\,mW (see ``Methods'' section) which consumes only 1.5$\%$ of the 1.5\,W cooling power at 3\,K. The chip comprises an array of Quantum Processor Units (QPUs) as shown in Fig.~\ref{fig:cryosys}(b). Each QPU houses a unique Quantum Core or the 2D qubit array. The qubit array is initialized by the reset transistors connected to the Quantum Point Contact (QPC) at the edges of the qubit array. This node is also shared with the detectors utilized during the read-out phase of the quantum experiment. The excitation signals for the qubits are sourced from digitally controlled injectors that are clocked by a high-speed pulse generator. The DC biasing of the qubit array is sourced from the Quantum Reference Bias Circuit (QRBC) coupled to the injectors \cite{bashir21}. Lastly, a pattern generator acts as a command and control block for all the aforementioned hardware interfacing the qubit array. The localized routing between the cryogenic electronics and the qubit array results in a power efficient control and detection system. 

The second unique feature is the nature of the position-based charge qubit. The quantum dots in the qubit array shown in Fig.~\ref{fig:cryosys}(b) are arranged over two rows in a ``double-V'' pattern~\cite{bashir2020single}. In each row, the quantum dots are isolated by a barrier controlled by an imposer. The QPC acts as a reservoir from which a single electron is tunneled to the first quantum dot. The qubits are designed using ``standard commercial'' CMOS process without any application-specific tailoring of the layer stack or materials. This means that the thin film of Si above the oxide~\cite{bashir2020single} is not a pure isotope. Additional nanoscopic impurities will be present at the $^{28}$Si/$^{28}$SiO$_2$ interface. The quantum information is encoded in electron's position controlled by a series of excitation signals applied at imposers to control the tunneling process of that electron across multiple quantum dots. The excitation signal can be a pulse~\cite{Gorman_2005} or resonant microwave signal~\cite{dkim15}. In a pulse driven mode, the width of the pulse controls the evolution of electron's wave function between the quantum dots while the DC biasing from the QRBC imposes a potential distribution across the wells \cite{Blokhina_2019}. In other words, the electron's position or occupancy across quantum dots varies over the duration of the pulse at a rate known as Rabi Frequency. In the microwave driven mode, the excitation frequency must be resonant with Larmor frequency, which is determined by the difference between the ground and excited energy levels, while no DC magnetic field is applied to the qubit array. Once the quantum operation concludes, the read-out phase begins. During the read-out phase, the final state of the qubit is measured by sensing the charge present on the QPC node.


The third unique feature is the switched-capacitor topology of the cryogenic electronics. The signals driving the imposers in the qubit array can be synthesized with simple and power-efficient circuits labeled as injectors in Fig.~\ref{fig:cryosys}(b). No active DC biasing is used in that block and the resulting transfer function is dependent primarily on the capacitors used in the design. Modeling of such circuitry can be done with reasonable accuracy as there is no dependency on transistor's $V_t$ or active currents which will vary significantly over temperature. The operating temperature is 100$\times$ lower than room temperature and consequently the $\frac{kT}{C}$ is lower. Furthermore, the experimental data (see  Sec.~\ref{sec:meas}) proves that the charge can be stored on a capacitor for 100's of $\mu$s due to the reduced leakage current at 3\,K.

The fourth distinction is the system apparatus shown in Fig.~\ref{fig:cryosys}(c).  There is no bulky passive RF circuitry or magnets. The qubit array is designed in a flip-chip package and soldered onto the DUT PCB such that the backside of the die is flush with the cold finger at 3\,K. The DUT PCB and flex cabling reside inside a heat shield with a penetration at the top for a two-stage Gifford-MacMahon (GM) cryocooler head. Helium-3 flows in the cryocooler through pipes connected between the compressor and the cryocooler head. The entire apparatus in Fig.~\ref{fig:cryosys}(c), except for the compressor sits, inside a 6-ft server rack.

The aforementioned distinctions can also serve as headwinds in the qubit performance. The first and foremost is the elevated temperature of the qubit array as a direct consequence of the single-chip integration and the thermal load specification of the cryocooler. Moreover, without a pure $^{28}$Si isotope, the decoherence process of this ``hot'' qubit array is influenced by the nanoscopic charge fluctuations caused by impurities. That said, characterizing the performance of charge and spin qubits under non-ideal conditions is an active area of research. Recently, Ramsey decay (T$_{\rm 2}$) of ~75\,ns and 200\,ns was measured for a spin qubit at 3.5\,K and 1\,K respectively~\cite{camenzind21}. Once performance trade-offs are understood and experimentally verified, the scientific community is likely to converge on a ``sweet spot" temperature setting that addresses the challenges in scaling and integration with an acceptable qubit fidelity. The architecture presented in this paper is a practical demonstration of a reasonable compromise between these critical parameters. The reduced decoherence times, as a result of this compromise, also imply that the qubits must be ultrafast in terms of gate flip times. Such architectures are well suited for hybrid classical-quantum algorithms that operate on low-depth circuits.


Next, we will detail each block within the Quantum Processor Unit.

\section{The Quantum Processor Unit (QPU)}
\label{sec:qc_app2}

\subsection{Quantum Core}
\label{sec:qcore}

\begin{figure}[t!]
     \centering
        \includegraphics[width=\textwidth]{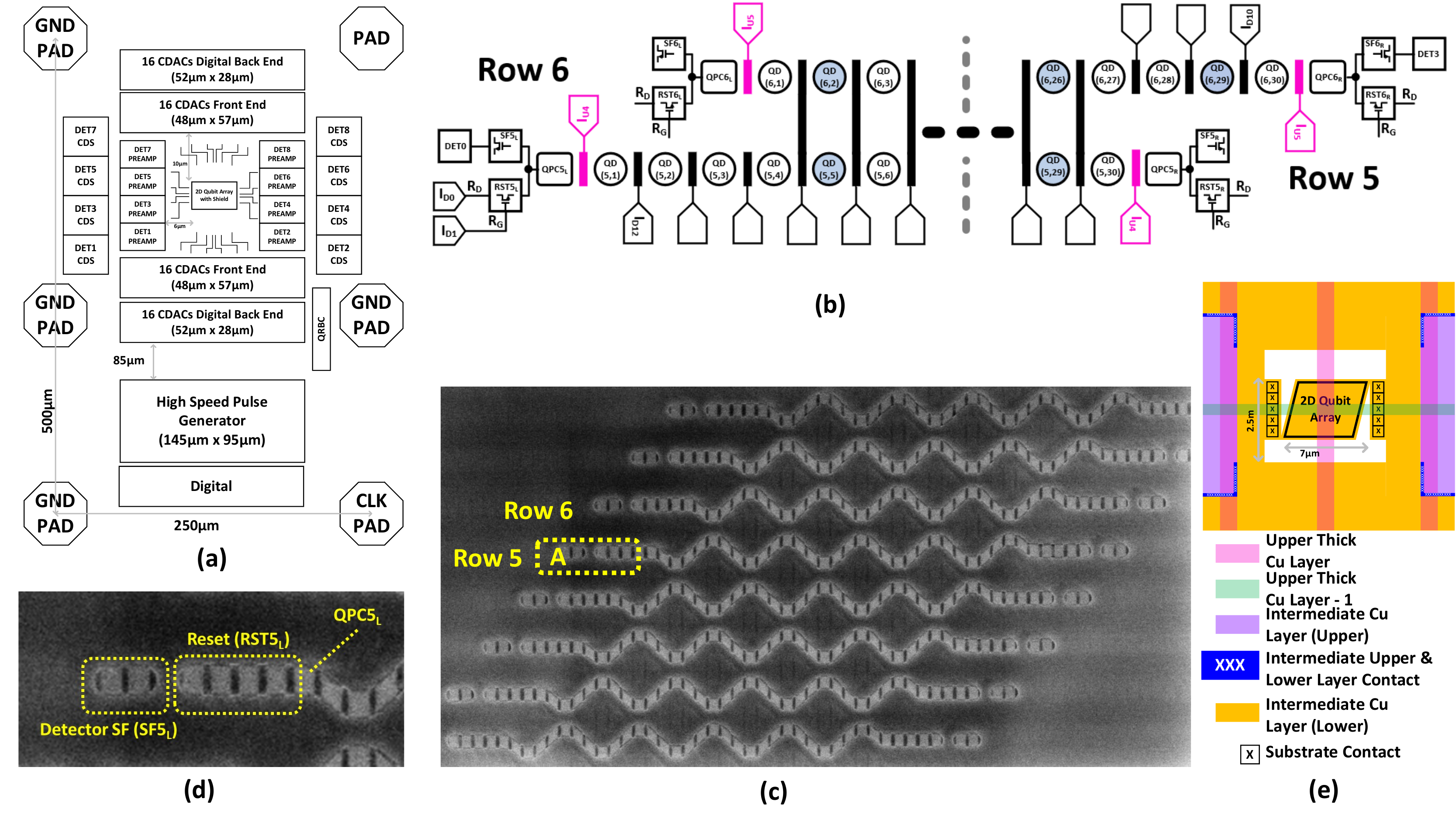}
        \caption{{\bf Details of Quantum Process Unit floor plan and the Quantum Core} (a) Quantum Process Unit floor plan. The injectors are placed above and below while the detectors are placed on the left and right side of the 2D qubit array. (b) Detailed drawing of Rows 5 and 6 in the 240 qubit array spanning. The coupled quantum dots are colored blue. (c) TEM photograph of the 240 qubit array. (d) Zoomed version of region 'A' in (c) showing the detector source follower and the reset transistor. (e) Layout design of the heat shield around the quantum structure.}
        \label{fig:qec201}
\end{figure}

The quantum processor unit contains the 2D 240-qubit array and all control electronics in a 250\,$\mu$m$\cross$500\,$\mu$m cell as shown in the floor plan in Fig.~\ref{fig:qec201}(a). 32 injectors and 8 detectors surround the quantum structure. All circuitry is at a distance of 10\,$\mu$m from the quantum core. The area of the quantum structure is 4\,$\mu$m$\cross$2\,$\mu$m, which is much smaller than the injector and detector circuits, and therefore, the wires from the quantum core have to fan-out to the surrounding circuits. The injectors comprise digital back-end and analog front-end circuitry. The analog front-end is an array of 16 capacitive DACs arranged in a tile~\cite{sscl20-esmailiyan}. The DAC input code is stored in registers residing in the digital back-end along with custom synthesized logic. The clocks for these CDACs are sourced from the high-speed pulse generator, which is located in the lower half of the cell and next to a 2\,GHz CLK pad. For the detector electronics, the CDS outputs are connected to a common row bus that is routed to the last stage of the detector chain. The map of the 240-qubit array and its TEM photograph is shown in Fig.~\ref{fig:qec201}(b) and (c), respectively. The 2D array is split into 8 rows and each row has 30 quantum dots. The notation of each quantum dot is $\rm QD(x,y)$ where ``x'' and ``y'' indicate the row number (range 1--8) and the position (range 1--30). The solid line between the quantum dots indicates an imposer. The edge of each row terminates into a quantum point contact ($\rm QPCX_{L,R}$) and the imposer (colored pink) between the $\rm QPCX_{L,R}$ node and the first quantum dot is controlled by CDACs labeled $\rm I_{Ux}$. The voltage from this CDAC controls the barrier and facilitates the transfer of a single electron from the QPC node to the first quantum dot. The remaining imposers control the tunneling of electron between neighboring quantum dots and perform Pauli rotations. The 8 rows in the 2D array are staggered, and when the line representing the imposer extends from one row to the next, it indicates that the imposers in those rows are controlled by a common CDAC. The sharing of CDACs is more frequent towards the middle of the array. The $\rm QPCX_{L,R}$ node is connected to reset and the source follower transistor as shown in the layout drawing in Fig.~\ref{fig:qec201}(d). Both of these devices are located close to the quantum structure in order to minimize the capacitance on QPC, which impacts the change in voltage as a result of single-electron transfer from this node. The reset transistor is a combination of four (4) stacked devices. A clearance of 10\,$\mu$m is used around the quantum structure to facilitate routing which is done in low-level metal layers. Additional thermal insulation comes from a metal shield placed on an intermediate-level copper metal layer above the quantum structure. That layer is tied to the substrate with contacts around the 2D qubit array as shown in Fig.~\ref{fig:qec201}(e). The shield is tied to the straps running horizontally and vertically on the thick copper layers, which connect with three ground pads located near the quantum structure.


\subsection{Quantum Reference Bias Generator}
\label{sec:vdac}

\begin{figure}[t!]
     \centering
        \includegraphics[width=\textwidth]{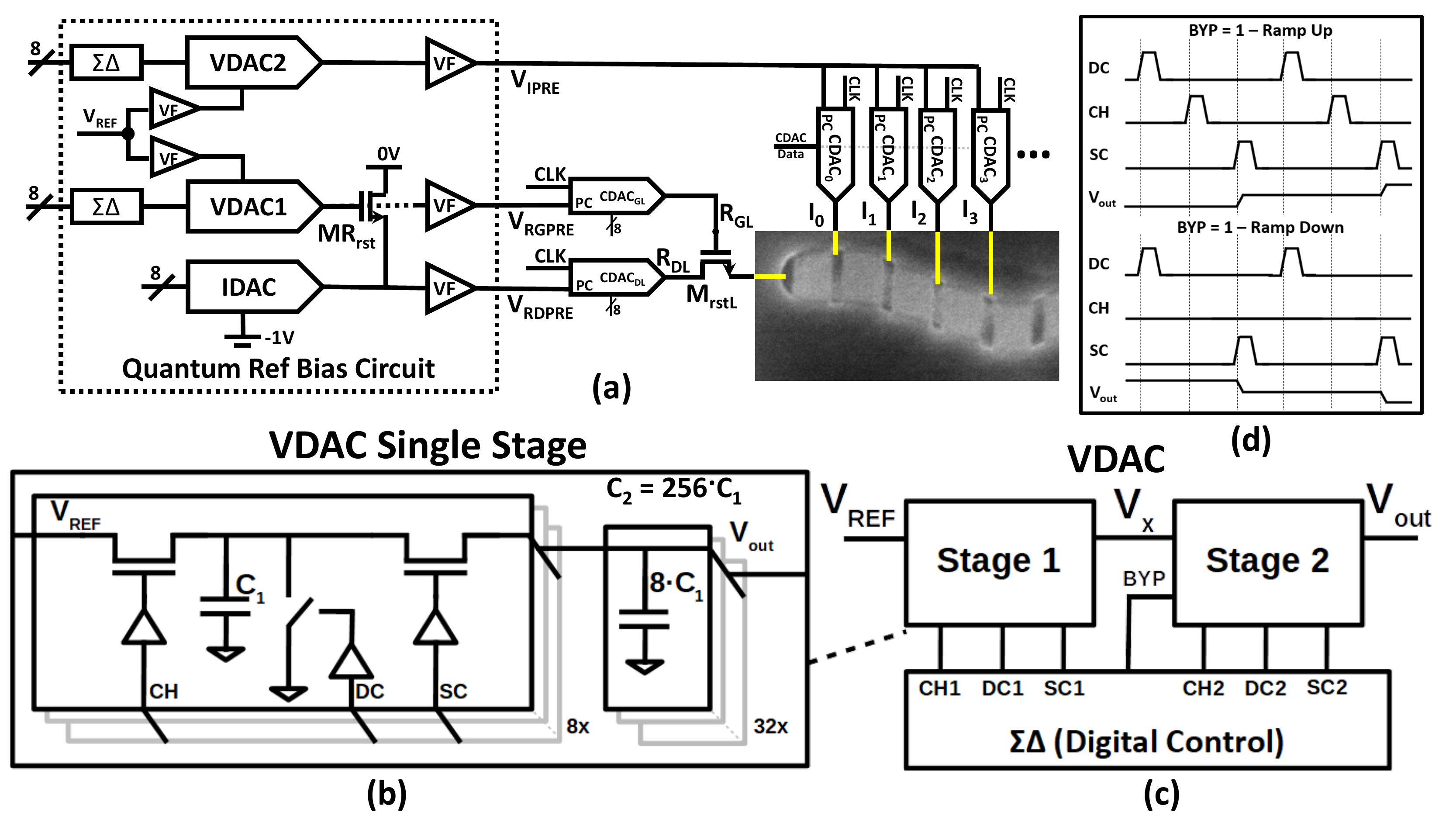}
        \caption{{\bf Quantum Reference Bias Circuit (QRBC).} (a) Top level block diagram. (b) Single VDAC stage. The charge is redistributed between $\rm C_1$ and $\rm C_2$. (c) Two cascaded VDAC stages. (d) VDAC signal sequence for output ramp up and down.}
        \label{fig:vdac}
\end{figure}

The quantum reference bias circuit is detailed in Fig.~\ref{fig:vdac}. The three critical nodes targeted for the dc biasing are the reset switch M$_{\rm rstL}$ drain RD, gate RG, and imposer gates I$\rm {_{U/Dx}}$ ($x = 1, 2, \cdots$). Their respective baseline voltages are V$\rm_{RDPRE}$, V$\rm_{RGPRE}$, and V$\rm_{IPRE}$. All these biases are sourced from a common reference V$_{\rm{REF}}$ that is external to the IC and has a long-term stability specification within $\pm$200\,ppm. V$_{\rm{REF}}$ is located on the control PCB outside the cryocooler. Two voltage followers (VF) drive the reference input V$_{\rm{REF}}$ to the VDAC1 and VDAC2 blocks. The VDAC blocks are switched-capacitor arrays that redistribute the charge between C$_1$ and C$_2$. Each VDAC comprises two cascaded stages for coarse and fine-tuning. The digital $\Sigma\Delta$ block generates pulses to discharge C$_1$ (DC), charge C$_1$ (CH), and transfer charge (SC) to C$_2$ based on an 8-b input. At the start, the DC, CH, and SC signals are pulsed in a sequence to ramp the output gradually towards the target. Once the output is around the target voltage, the CH pulse activity decreases. At this point, the resolution of the VDAC is approximately 300\,$\mu$V given the capacitor ratio. Then, the second stage can be activated for finer adjustments to hit the target value. Next, the clock activity ceases, and the VDAC output is frozen before the CDAC pre-charge (PC) switch is enabled just before the quantum experiment commences. From this point onwards, any adjustment above or below the pre-charge level applied to the qubit array can be dynamically shifted by the CDACs based on their unique 8-bit digital control. V$\rm_{RDPRE}$ is generated by V$\rm_{gs}$ of a replica of the M$\rm_{pre}$ reset switch, MR$\rm_{pre}$, and it is further adjusted by the 8-bit current DAC (IDAC) with a range from a few nA to 1\,$\mu$A (max). In order to generate a negative V$\rm_{RDPRE}$, IDAC V$_{\rm SS}$ is set to $-$1\,V while MR$\rm_{pre}$ drain is at 0\,V.

Two environmental factors work as tailwinds in this design. One is the reduced leakage current at cryogenic temperature (see Sec.~\ref{sec:meas}). This implies that C$_1$ and C$_2$ can hold their charge long enough in cryogenic conditions for a quantum experiment to conclude, which typically takes 100's of ns. The other aiding factor is the lower $kT/C$ noise at 3\,K, which is 1/100th compared to the room temperature. Therefore, the capacitor size can be scaled down resulting in a miniaturized layout. The three internal voltage followers (VF) at the outputs shown in Fig.\,\ref{fig:vdac} are only needed when driving external pads for testing purposes and can be eliminated. All capacitors in the circuit are MOMCAPs designed with a stack of 3--5 intermediate metal layers. The circuits are designed with nominal $V_{\rm th}$ (RVT) transistors without any body biasing applied while the main supply is 0.8\,V. The bias voltages applied to the quantum core have a maximum limit of 0.8\,V-$V\rm{_{th}}$ and a minimum limit of -1\,V+$V\rm{_{dsat}}$, which is a sufficient range in order to conduct quantum experiments on the QDA. The typical clock frequency of the $\Sigma\Delta$ is 125\,MHz and the switches in VDACs are sized appropriately to charge and discharge capacitor C$_1$ during the pulse duration of 8\,ns. This gives reasonable dynamic power dissipation given the specified clock frequency. At the specified rate, VDAC outputs stabilize within 100\,$\mu$s after activation.
 
\subsection{Injector}
\label{sec:cdac}

\begin{figure}[t!]
     \centering
        \includegraphics[width=\textwidth]{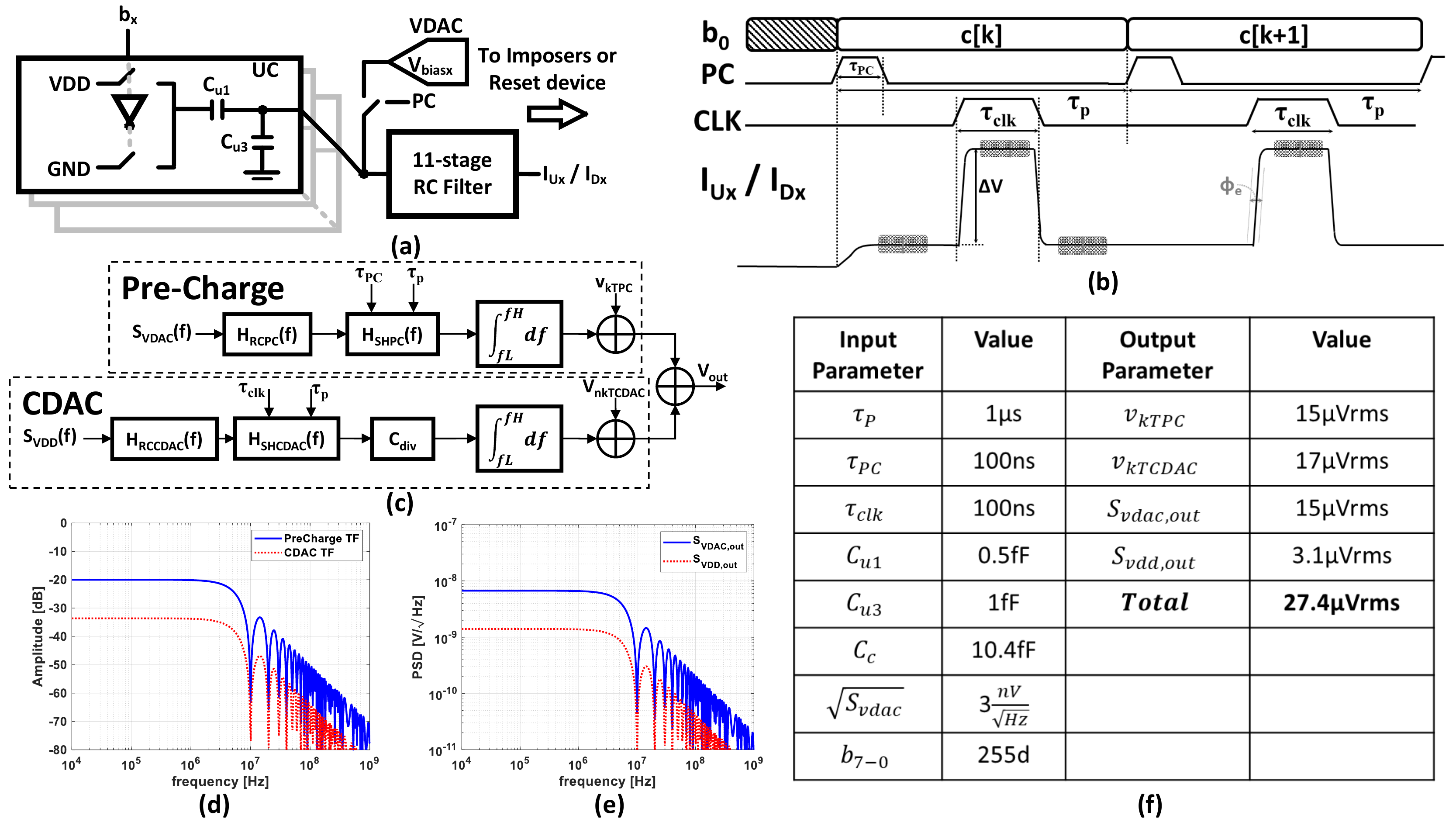}
        \caption{{\bf Injector functionality and noise characteristics} (a) Block level diagram of the injector circuit. (b) Time-domain waveform of the injector output. The gray mesh region on the pedestal and step voltage represents noise. (c) Injector noise analysis with supply and VDAC noise at inputs. The noise is injected from the pre-charge and CDAC switch capacitor path and is summed together at the final output. (d) Amplitude response of transfer function of Pre-charge and CDAC path. (e) The VDAC and VDD noise spectra at the injector output. The noise from pre-charge path dominates the cumulative noise. (f) Noise model parameters and output noise summary.}
        \label{fig:cdac}
\end{figure}

The principal function of the injector~\cite{sscl20-esmailiyan} is to manipulate the reset device before and during the quantum experiment, control the barrier at the single-electron injection device, and to excite the imposers to control the evolution of electron's wave function in the intermediate quantum dots. The block diagram shown in Fig.~\ref{fig:cdac}(a) illustrates the injector function. First, the imposer node labeled $\rm I_{Ux}$ or $\rm I_{Dx}$ is pre-charged to a bias voltage $\rm V_{biasX}$ which is sourced from VDAC output inside the Quantum Reference Bias generator. This sets the pedestal of the voltage pulse shown in Fig.~\ref{fig:cdac}(b). Next, the digital code $\rm b_X$ activates the switch capacitor array, and a charge transfer takes place between $\rm C_{u1}$ and $\rm C_{u3}$. This elevates the injector output by $\Delta \rm V$, which is digitally controlled. The 11-stage RC filter has a pole beyond 100\,GHz and filters any ultra-fast transients on the imposer node. 

The transfer function analysis shown in Fig.~\ref{fig:cdac}(c) considers the noise contribution from two paths: the pre-charge path and the switch capacitor array of the CDAC path. $S_{\rm VDD}$ and $S_{\rm VDAC}$ are the PSD of noise voltage on the $V_{\rm DD}$ supply and VDAC output, respectively. The transfer functions $H_{\rm RCPC}$ and $H_{\rm RCCDAC}$ model the RC filter formed by the switch resistance and output node capacitance. $H_{\rm SHPC}$ and $H_{\rm SHCDAC}$ model the aliasing of input noise due to the sample and hold operation\cite{fischer1982jssc} and is impacted by the timing parameters $\tau_{\rm p}$, $\rm \tau_{PC}$, and $\tau_{\rm clk}$. The formulas for these two transfer functions are derived in the ``Methods'' section. $\rm C_{div}$ is a scaling factor due to capacitive divider formed by $\rm C_{u1}$ and $\rm C_{u3}$. The noise spectra in the pre-charge and CDAC paths are integrated over frequency and the resulting noise power is added to $\frac{kT}{C}$ noise ($v_{\rm kTPC}$ and $v_{\rm kTCDAC}$) to compute the total noise at the injector output. Figure~\ref{fig:cdac}(d) compares the amplitude response in the ``pre-charge'' and the CDAC paths. The gain of the CDAC path is attenuated by $\rm C_{div}$ and, therefore, the VDAC noise at the injector output dominates the supply noise as shown in Fig.~\ref{fig:cdac}(e). Based on the circuit parameters defined in Fig.~\ref{fig:cdac}(f), the cumulative noise at the injector output is 27.4\,$\mu$V-rms.


\subsection{Detector}
\label{sec:det}

\begin{figure}[t!]
     \centering
        \includegraphics[width=\textwidth]{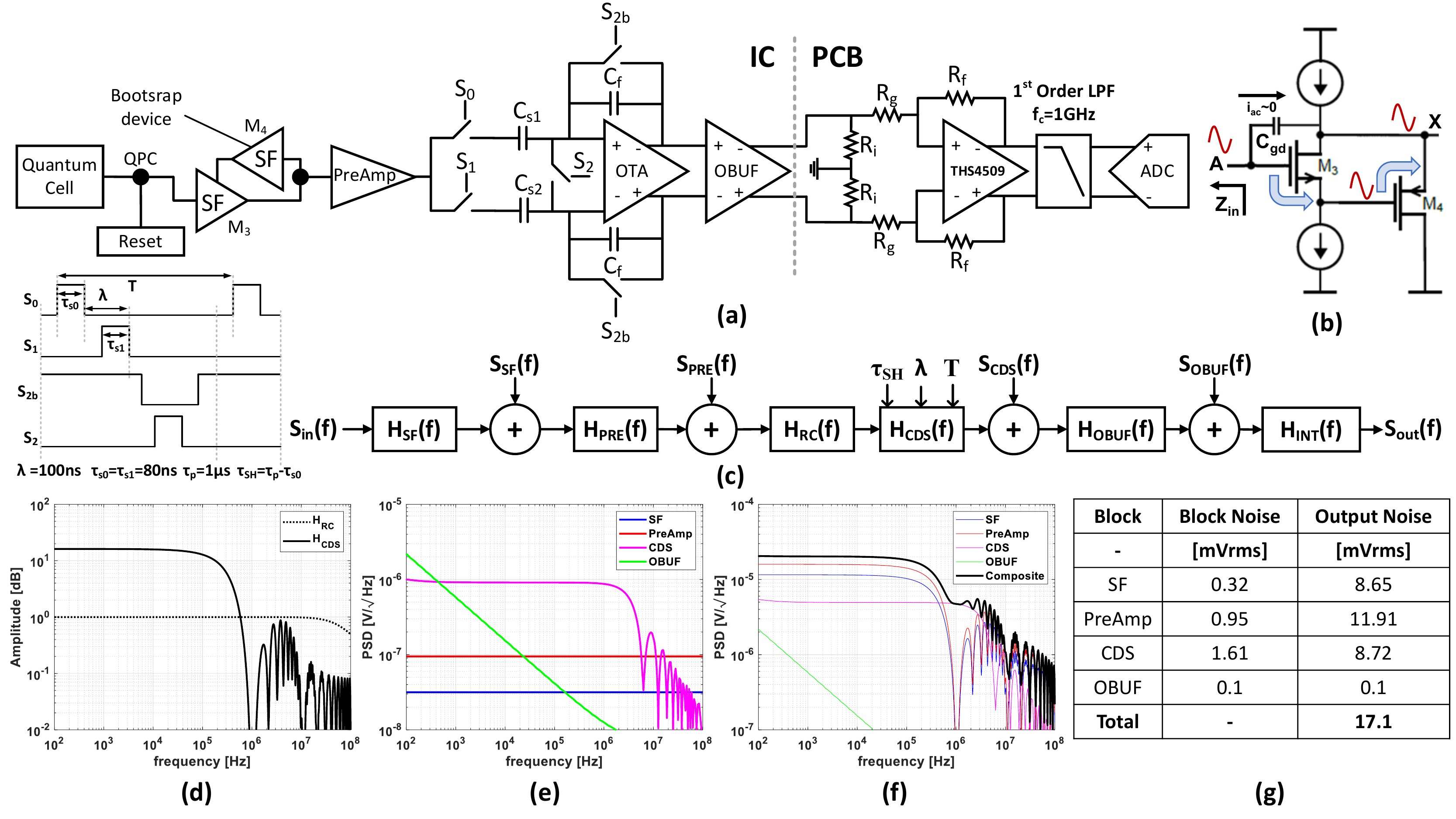}
        \caption{{\bf Detector chain noise sources and transfer function.} (a) Block level diagram of detector chain. (b) $\rm C_{gd}$ neutralization using bootstrap device M4. (c) Frequency domain model of the detector chain. (d) $\rm H_{CDS}$ transfer function. (e) Spectra of noise sources used in the model. (f) Noise contribution from all sources at the detector output. (g) Summary of noise analysis.}
        \label{fig:det}
\end{figure}

The detector chain, shown in Fig.~\ref{fig:det}(a) \cite{sscl20-esmailiyan}, comprises the following stages: source follower, pre-amplifier, CDS (correlated double sampler), output buffer (OBUF), interface circuit with LPF, and ADC. The interface circuit and the ADC are off-chip components. The detector has to interface with a high-impedance node labeled QPC which is connected to the quantum structure and a reset device. A detector input impedance, which is substantially higher than the sourcing node (i.e. QPC) ensures a low loss transmission of the signal from the source node into the detector. For this reason, the source follower transistor $\rm M_3$ in Fig.~\ref{fig:det}(b) is designed with minimum dimensions. However, any parasitic capacitance such as $C_{\rm gd}$ arising from $\rm M_3$ can limit the input impedance of the detector. To address this issue, a bootstrapping device in the form of a source follower~\cite{chi2011jcas} $\rm M_4$ is added to the circuit as shown in Fig.~\ref{fig:det}(b). This device creates a positive feedback at node A and forces the voltage at node A and X to move in tandem as a result of the perturbation caused by movement of charge from QPC during a quantum experiment. As a result, the current through $C_{\rm gd}$ and hence the impact of this capacitance is minimized.

Next, the transfer function of the detector chain will be analyzed. Figure~\ref{fig:det}(c) shows the transfer function from the QPC node till the ADC input. The source follower $\rm H_{SF}$, the pre-amplifier $\rm H_{PRE}$, and the output buffer $\rm H_{OBUF}$ transfer functions are assumed to be constant over frequency with a gain of 0.9, 2.2, and 5.4, respectively. $H_{\rm RC}$ is the RC response due to switch resistance and sample capacitor ($C_{\rm sx}$) in the CDS, which has a pole around 70-MHz. $H_{\rm CDS}$ is the CDS transfer function which also accounts for aliasing of input noise due to the sample-and-hold operation and is impacted by three parameters shown in the timing diagram of Fig.~\ref{fig:det}(c): $\tau_{\rm SH}$, $\lambda$, and $T$. The derivation of this transfer function is based on \cite{fischer1982jssc,pimbley1991tcas} and will be described in detailed in the ``Methods'' section. Figure~\ref{fig:det}(d) shows the $H_{\rm CDS}$ amplitude response with $\tau_{\rm S0,S1}$, $\lambda$, and $T$ set to 80\,ns, 100\,ns, and 1\,$\mu$s, respectively. $H_{\rm INT}$ models the ADC interface designed using a THS4509 op-amp and a low-pass filter. The resistor values for $R_f$ and $R_g$ are chosen such that the gain of the interface is 2. $R_i$ sets the load line for the OBUF circuit, which is operating in class-A.   

The spectra of noise sources in the detector are shown in Fig.~\ref{fig:det}(e). The pre-amplifier and the source follower noise is modeled as a white noise source in order to simplify the analysis of noise shaping due to $H_{\rm CDS}$ (see ``Methods'' section). The contribution of each noise source at the output is computed by multiplying the PSD of noise voltage by its respective transfer function. The results of that exercise are summarized in Fig.~\ref{fig:det}(f) and (g). The noise at close-in frequencies is dominated by the pre-amplifier and source follower due to the aliasing operation in $H_{\rm CDS}$. The composite integrated noise at the detector output is 17\,mV-rms, which is in line with the measurement reported in \cite{sscl20-esmailiyan}. It is important to note that the information from the quantum experiment that is stored in the capacitors $C_{\rm sx}$ is band-limited by the RC network modeled by $H_{\rm RC}$.



\subsection{Pattern Generator (PATGEN)}
\label{sec:patgen}

\begin{figure}[t!]
     \centering
        \includegraphics[width=\textwidth]{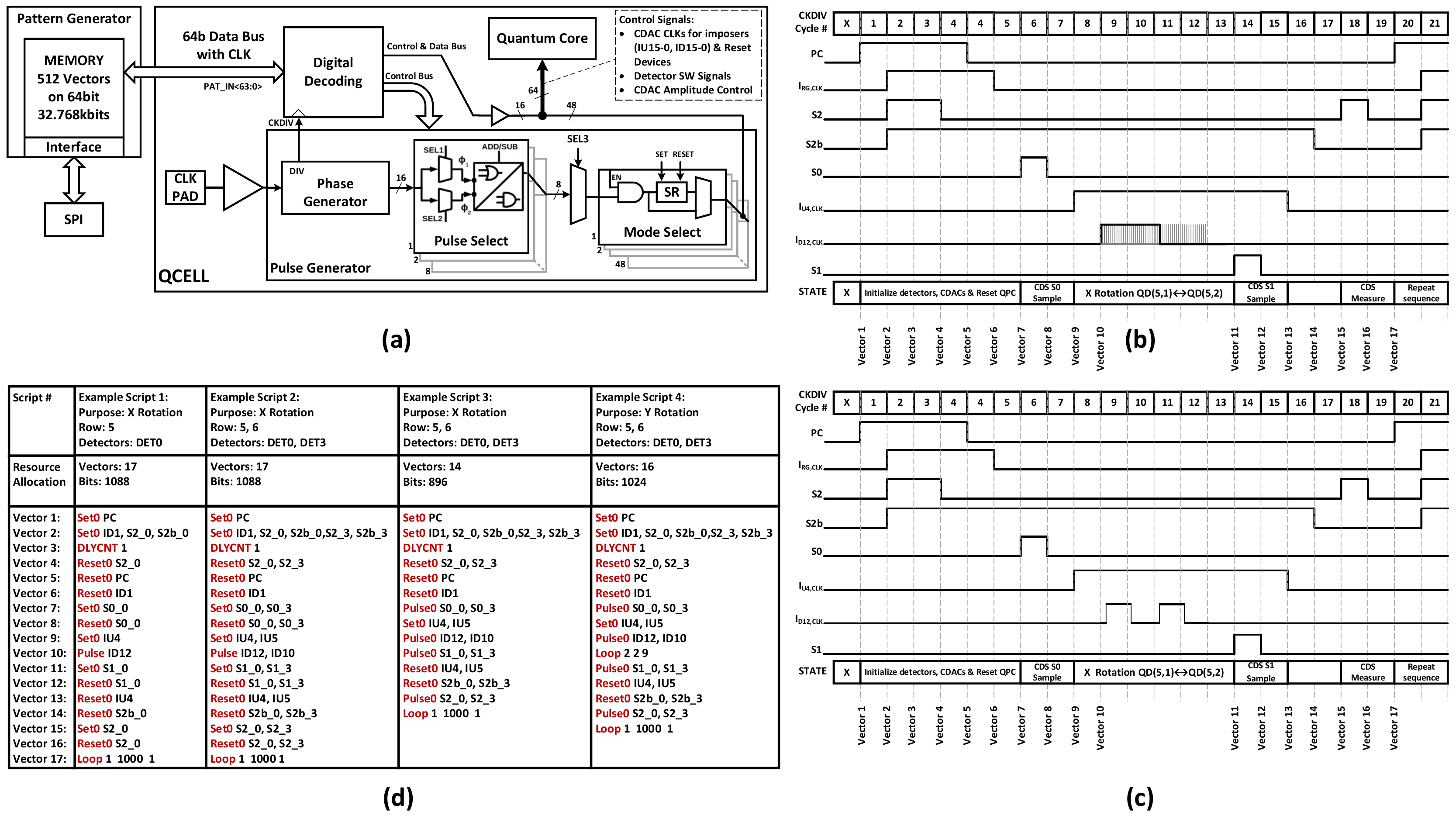}
        \caption{{\bf Pattern Generator, its data path, and four examples of resource allocation of pattern memory.} (a) Data path flow from PATGEN output to the Quantum Core. The control signals from the pattern drive the Pulse Generator to synthesize precisely timed signals into the Quantum Core. (b) Signal timing of Example Script 1 in which an X Pauli rotation is performed using a pulsed excitation~\cite{Gorman_2005}. (c) Signal timing of Example Script 4 in which an X Pauli rotation is performed using resonant driving~\cite{dkim15}. (d) Summary table of four quantum experiment test scripts and their resource allocation.}
        \label{fig:patgen}
\end{figure}

The pattern generator (PATGEN), shown in Fig.~\ref{fig:patgen}(a), acts as a command and control block in the Quantum Processor Unit and is integrated with the extensive electronics and the qubit array. The PATGEN comprises a memory core, an interface unit, and decoding logic. The interface unit handles the signaling between the 4-wire SPI and the PATGEN. The 32-kbit memory core has a depth of 512 while the data output bus is 64-bit wide. The memory core can store 512 vectors and a vector corresponds to a timing event caused by an instruction in the programming script. This will be explained in detail shortly. Inside the QCELL block, the 2\,GHz clock is divided to generate CKDIV. The desired pattern is loaded into the memory core and channeled onto the 64-bit data bus while the timing of the pattern is based on CKDIV clock. The digital decoding block deciphers the incoming data and routes the information to two output ports: the control bus for the high-speed pulse generator~\cite{esscirc_2019} and the control and data bus for the Quantum Core. The 16 bit data bus for the quantum core contains the  8-bit CDAC amplitude word and some control signals. The remaining 48-bit control lines are clock signals for the 32\,CDACs and control signals (CDS switches) for the eight (8) detectors. The clock and control signals are generated by the pulse generator and that process is described next.

The first stage of the pulse generator is a Phase Generator designed using a 16-stage Johnson counter that generates multiple signals with a phase separation of 500\,ps for a 2\,GHz clock. The clock frequency also sets the timing resolution of the pulse generator. Two signals from the Johnson counter are chosen for a programmable AND/OR operation, which either narrows or expands a pulse in the Pulse Select block. Up to 8 of these blocks run in parallel while each leaf cell has its unique set of SEL1, SEL2, and AND/OR selection codes in order to generate multiple signals with varying pulse-widths at the same time. This is a useful feature when there is a need to switch between two different timing profiles seamlessly. Finally, one output is selected to be passed to the Mode Select block which either propagates the input to the output or passes it to a SR (set/reset) latch. The latter option allows the pulse duration to be extended over longer time scales.

After describing the hardware implementation, it is important to discuss the programming and resource allocation per quantum operation or, more specifically, an X Pauli rotation. The nature of the Pauli rotation exhibited by the charge qubit is a function of energy levels and well potential over the qubit array in addition to the waveform applied to the imposer~\cite{giounanlis2021cmos}. The pattern shown in Fig.~\ref{fig:patgen}(b) and its corresponding program script in Fig.~\ref{fig:patgen}(d) (Example Script 1) is an example of a pulse-driven X rotation. This example has 17 vectors or instructions and each is annotated by a colored font. The number proceeding the instruction links that instruction with the Pulse Select block number. For instance, the $\rm Set0$ and $\rm Set1$ commands are generated by Pulse Select Leaf Cell no. 1 and 2 out of the eight leaf cells. $\rm Pulse0$ generates a rising and falling edge based on the 16 phases of the Johnson counter output. This is very useful in quantum experiments where fine timing resolution is required. The $\rm Loop$ command at Vector 17 allows the sequence to be repeated 1000 times (2nd parameter) while the 3rd parameter (1 in this case) indicates the Vector number to start the loop from. The total memory allocation for this script is 1088 bits (17$\cross$64b) and the quantum operation is only isolated to quantum dots $\rm QD(5,1)$ and $\rm QD(5,2)$ in Row 5 of the qubit array (coupled to DET0) as shown in Fig.~\ref{fig:qec201}(b) and (c).

\begin{figure}[t!]
     \centering
        \includegraphics[width=\textwidth]{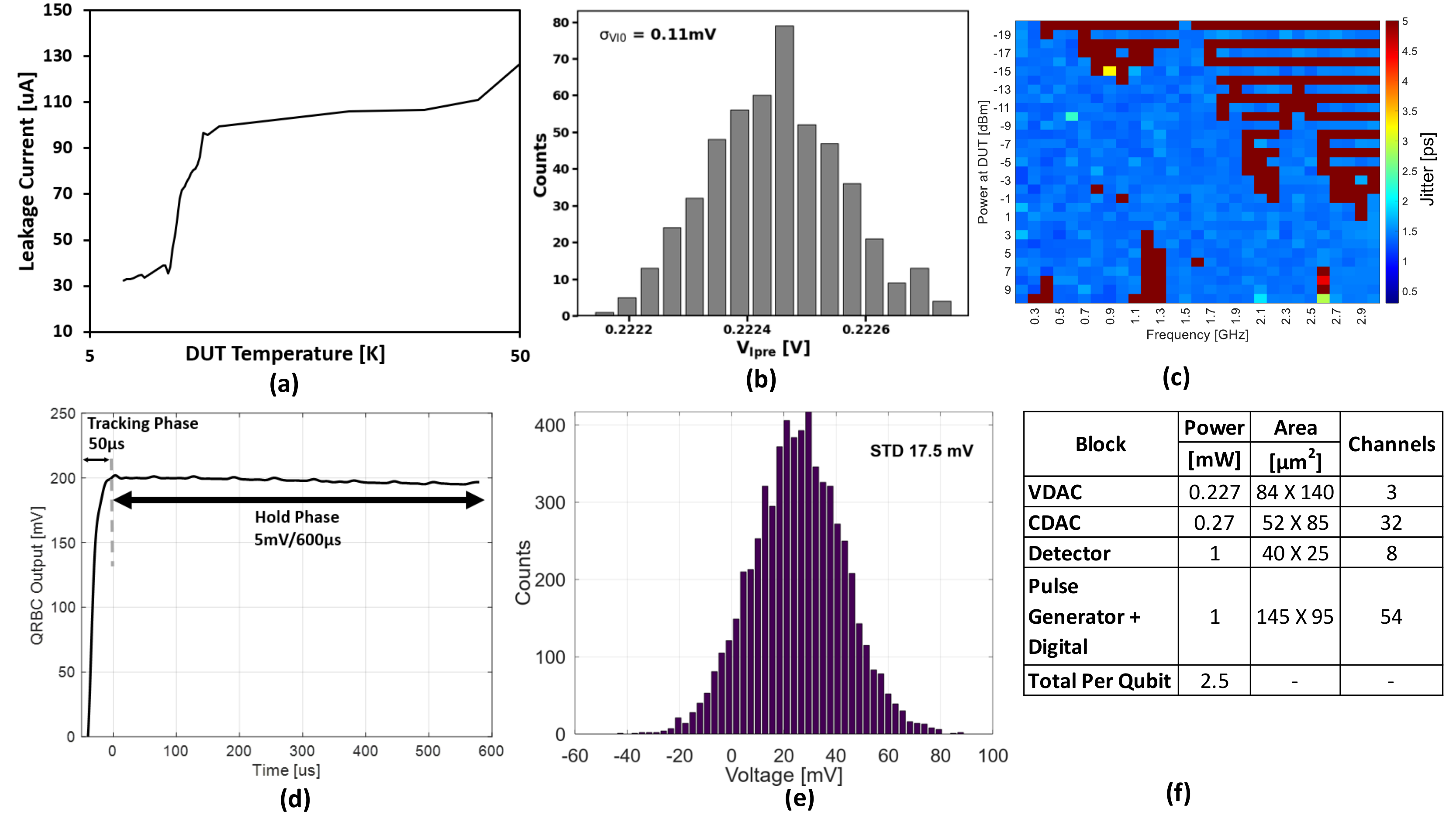}
        \caption{{\bf IC Measurements.} (a) Total IC leakage current over temperature. (b) Noise at Quantum Reference Bias Circuit (QRBC) output connected to the imposer. (c) Heat map of RF clock jitter over signal generator frequency and power. (d) QRBC output over long time span. The output droop is negligible over 100\,$\mu$s and 5\,mV over 600\,$\mu$s. (e) Detector output noise measured at 3K. Standard deviation is 17.5\,mV. (f) Summary of power of the entire control and detection circuits including the digital blocks such as the pulse generator and the PATGEN. The total power budget is 2.5\,mW per qubit.}
        \label{fig:meas1}
\end{figure}

In the next step, the benefit of the parallel bus operation of the pattern generator is realized in Example Script 2 (Fig.~\ref{fig:patgen}(d) third column) where multiple node names are added to the same Vector. The same X rotation script now operates on four quantum dots over two rows and two detectors (DET0 and DET3); however, the memory utilization remains unchanged. Example Script 3 is a more area efficient rendering of the aforementioned script where the memory utilization is down to 896 bits. Example Script 4 in the last column column in Fig.~\ref{fig:patgen}(d) performs a X Rotation using a resonant driving method~\cite{dkim15} which is facilitated by the nested loops. The nested loop is at Vector 11 and loops two times starting from the position of Vector 9. This useful technique results in a memory allocation of 1024 bits, which is not too far from Example Script 3. In summary, the key feature in this design is a memory efficient instruction set operating on the pattern generator to create signals on the 64-bit bus which enables highly versatile operation on a large qubit array, such as the one presented in this work.

\section{Measurement Results}
\label{sec:meas}

The measurements presented in this section are divided into classical electronic measurements and quantum experiment measurements. Fig.~\ref{fig:meas1} underscores the benefits arising from the low-temperature environment that acts as tailwind in cryogenic control circuit design. One of these is the low leakage current as demonstrated in Fig.~\ref{fig:meas1}(a) where the total leakage current of the entire IC goes down as the DUT temperature is reduced to 6\,K. The low leakage allows the capacitor to hold its charge over a long period of time as shown in the time-domain quantum reference bias circuit (QRBC) output in Fig.~\ref{fig:meas1}(d). The QRBC output stabilizes in 50\,$\mu$s and it remains so for nearly 100\,$\mu$s during the hold phase. The long-term droop in the QRBC output is only 5\,mV in 600\,$\mu$s. The QRBC output driving the imposer is measured over thousands of trials, and the output distribution is plotted in Fig.~\ref{fig:meas1}(b). The standard deviation of this distribution is 110\,$\mu$V which is lower than the $\frac{kT}{q}$ limit of 267\,$\mu$V at 3\,K. Fig.~\ref{fig:meas1}(c) demonstrates the robustness of the high-speed pulse generator, which relies on an off-chip coax cable and frequency tuning elements~\cite{Bashir20rfic}. The clock signal path should be low loss over a wide bandwidth, which enhances the operating range and resolution of the pulse generator. Low level of mismatch implies that the DUT can operate at low output power from the signal generator, reducing the amount of active heat dissipation. The heatmap shown in Fig.~\ref{fig:meas1}(c) indicates a large range of frequencies and signal powers at the DUT input where the clock jitter is around 1.5\,ps. This performance is achieved at the highest frequency of 2.5\,GHz at $-$10\,dBm delivered into the DUT. Fig.~\ref{fig:meas1}(e) is the detector noise measurement which is in line with the theoretical calculations detailed in Sec.~\ref{sec:det}. The total power consumption of all the blocks required to operate on one qubit is summarized in Fig.~\ref{fig:meas1}(f). 

\begin{figure}[t!]
     \centering
        \includegraphics[width=\textwidth]{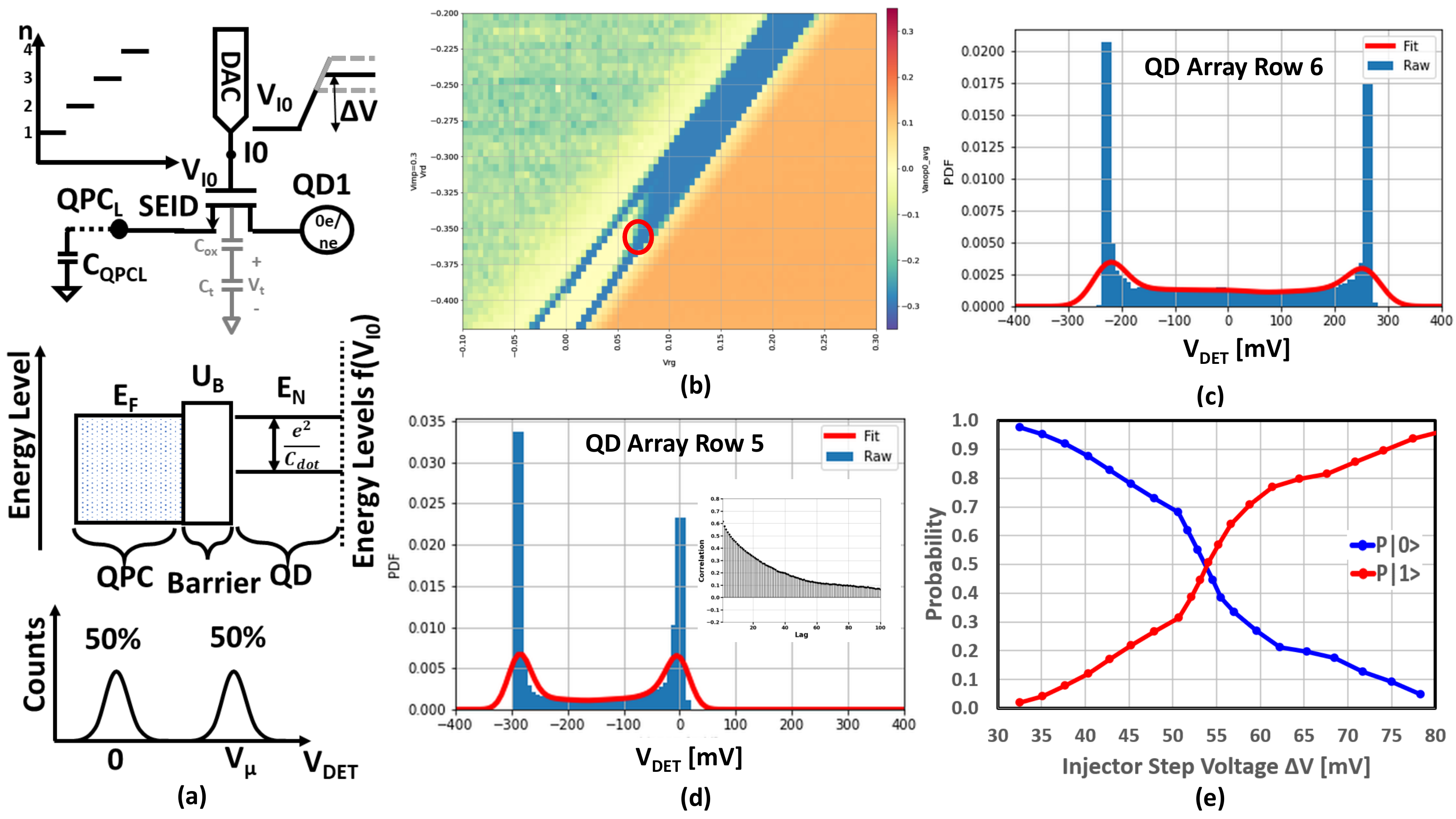}
        \caption{{\bf Experimental demonstration of Coulomb blockade at the SEID device on two rows of the 2D qubit array in Fig.~\ref{fig:qec201}(b).} (a) Graphical representation of the condition required for quantum resonant tunneling. (b) Detector average voltage heat map as a function of reset transistor bias $\rm V_{RD}$ and $\rm V_{RG}$. The region of interest is highlighted as the red circle where the detector output shows bi-modal signature. (c) The bi-model detector output on DET3 located on 6th row in Fig.~\ref{fig:qec201}(b). (d) The bi-model detector output on DET0 located on 5th row in Fig.~\ref{fig:qec201}(b). The figure inlet shows the auto-correlation function and the response highlights a key characteristic of quantum resonant tunneling process~\cite{li_z_scirep18,li_z_iedm17,h_miki12}. (e) $\rm P|0\text{\small >}$ and $\rm P|1\text{\small >}$ probability as a control function of injector step voltage similar to the gate modulation of probability in ref.~\cite{li_z_scirep18,li_z_iedm17,h_miki12}.}
        \label{fig:meas2}
\end{figure}

Fig.~\ref{fig:meas2} details the quantum measurements performed on quantum dots located on rows 5 and 6 (coupled to DET0 and DET3) respectively of the 2D qubit array in Fig.~\ref{fig:qec201}(b). A sequence similar to the one shown in Fig.~\ref{fig:patgen}(b) is applied. The QPC node can be initialized by first scanning voltages on the reset transistor gate ($\rm R_G$) and drain ($\rm R_D$). The imposer connected to the single-electron injection device (SEID) is pulsed with a set amplitude during this scan. The detector output is stored over many iterations and the average value is plotted as a heat map in Fig.~\ref{fig:meas2}(b). In this plot, the region highlighted by the red circle is of particular interest where the detector output shows a bi-model behavior. At this bias point, the conditions for quantum resonant tunneling depicted in Fig.~\ref{fig:meas2}(a) are met. The detector output distribution shows two distinct peaks when the barrier with potential energy $\rm U_B$ is low while the Fermi energy level $\rm E_f$ at QPC aligns with a discrete energy level in the adjacent quantum dot $\rm E_N$. The outcome of each experiment is one of two possibilities. Either the electron tunnels to the adjacent quantum dot resulting in a loss of charge at QPC sensed by the detector resulting in a output voltage $\rm V_{\mu}$. The probability of this event is $\rm P|1\text{\small >}$. In the other event, the electron does not tunnel resulting in a detector output at 0\,V. The probability of this event is $\rm P|0\text{\small >}$. Figures~\ref{fig:meas2}(c) and (d) show the histograms of detectors DET3 and DET0 respectively when the reset transistor is biased properly and the events associated with the two states, namely $\rm |0\text{\small >}$ and $\rm |1\text{\small >}$, are clearly visible with a separation of 300\,mV between them at the detector output. The quantum resonant tunneling process is moderately correlated when the lag is short as shown in Fig.~5 in ref.~\cite{li_z_scirep18}. This unique characteristic is also observed by performing auto-correlation on the time series data of the detector samples and the result is shown in Fig.~\ref{fig:meas2}(d) inlet. $\rm P|1\text{\small >}$ increases with increasing step amplitude of the SEID injector (or CDAC code) while $\rm P|0\text{\small >}$ retrogrades as shown in Fig.~\ref{fig:meas2}(e). This is yet another characteristic of quantum resonant tunneling also shown in Fig. 3 of ref.~\cite{li_z_scirep18} and Fig. 9 of ref.~\cite{li_z_iedm17}. The x-axis in this chart has been translated from CDAC code a step voltage (based on Fig.8(a) in ref.~\cite{sscl20-esmailiyan}). Therefore, we infer that in this state, the electron injection process is precisely controlled by a digitally controlled circuit parameter allowing us to manipulate the quantum information in two quantum dots (Row 5 and 6) in the 2D qubit array.

\begin{figure}[t!]
     \centering
        \includegraphics[width=\textwidth]{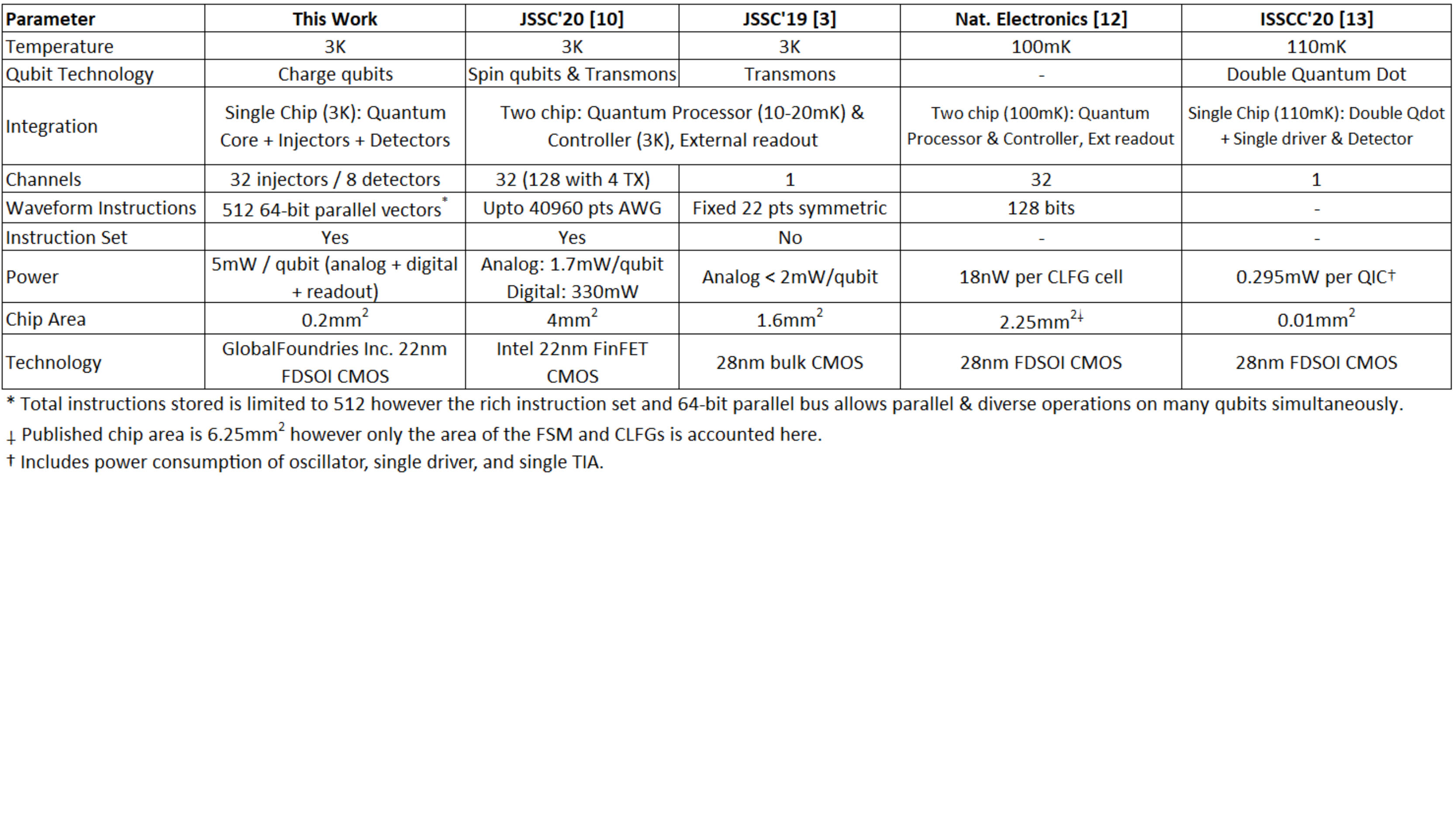}
        \vspace{-45mm}
        \caption{{\bf Comparison with prior art.}}
        \label{fig:comp}
\end{figure}

\section{Conclusions}

We have presented a monolithic integration of a quantum resonant tunneling gate with a quantum controller and read-out system. The entire apparatus operates on a low power budget of 2.5\,mW per qubit. This figure of merit will improve further with on-chip digitization of detector output and reduce the cabling to room temperature electronics down to 13 wires. The monolithic integration is performed in ``standard commercial'' CMOS technology. The power efficient architecture of cryogenic electronics is enabled by switch capacitor implementation which is able to hold charge well over the time period required for a quantum experiment. The measurement results demonstrate a precise control of coulomb blockade and the electron injection process in two rows of the 240 qubit array. The total thermal load (calculated in ``Methods'' section) with current apparatus is 1.5$\%$ of the available cooling power at 3\,K and still remains well below the 1.5\,W limit when the  2D qubit array is scaled to 2400 qubits.

\section*{Methods}
\subsection*{Experimental Data Processing}
\label{sec:data_proc}

This section details important calculations associated with Fig.~\ref{fig:meas2}(c,d,e). The difference in $\rm V_{DET}$ between $\rm |0\text{\small >}$ and $\rm |1\text{\small >}$ of 300\,mV can be translated to voltage fluctuation of 3.75\,mV based on a simulated gain of the detector chain of 80\,V/V~\cite{sscl20-esmailiyan}. This is a factor of 12x higher than $\frac{kT}{q}$ limit of 300\,$\mu$V and gives us a reasonable SNR at the detector input. However, the fluctuation of the signal at the QPC node due to resonant tunneling process is wide-band and can't be processed by the detector. Consequently, there are counts appearing in the middle zone i.e. $\rm V_{DET}$ values between -290\,mV and -10\,mV in Fig.~\ref{fig:meas2}(c,d). Those events have been removed when analyzing probabilities shown in Fig.~\ref{fig:meas2}(e). The total counts for $\rm P|0\text{\small >}$ is determined by adding events in three bins in Fig.~\ref{fig:meas2}(c,d): -300\,mV bin and two adjacent bins. Similar process is applied to compute total counts for $\rm P|0\text{\small >}$ around 0\,mV bin. The probability is simply the ratio of the event counts and the total sample space (10k). The probabilities are plotted for a range of injector step voltages in Fig.~\ref{fig:meas2}(e).

Next, we estimate the difference in energy level $\rm \Delta E$ which is given by $\rm \frac{e^{2}}{C_{dot}}$ where $\rm e$ is the charge of an electron. In Fig.~\ref{fig:meas2}(e), $\rm P|1\text{\small >}$ fully transitions from 0$\%$ to 95$\%$ when the step voltage of the injector is increased from 33\,mV to 78\,mV. This 45\,mV difference corresponds to a $\Delta V_t$ of 4.5\,mV at the tunnel junction due to capacitive division from $\rm C_{ox}$ and $\rm C_t$ (see Fig.~\ref{fig:meas2}(a)). From this parameter, we calculate $\rm \Delta E$ to be 4.5\,meV and $\rm C_{dot}$ to be 35\,aF. $\rm \Delta E$ is 10x above the $\rm kT$ noise of 0.27meV at 3\,K and $\rm C_{dot}$ value seems reasonable for a quantum dot of size 80\,nm$\cross$80\,nm.



\subsection*{Heat Load Calculations}
\label{sec:heat_load}

\begin{figure}[t!]
     \centering
        \includegraphics[width=0.7\textwidth]{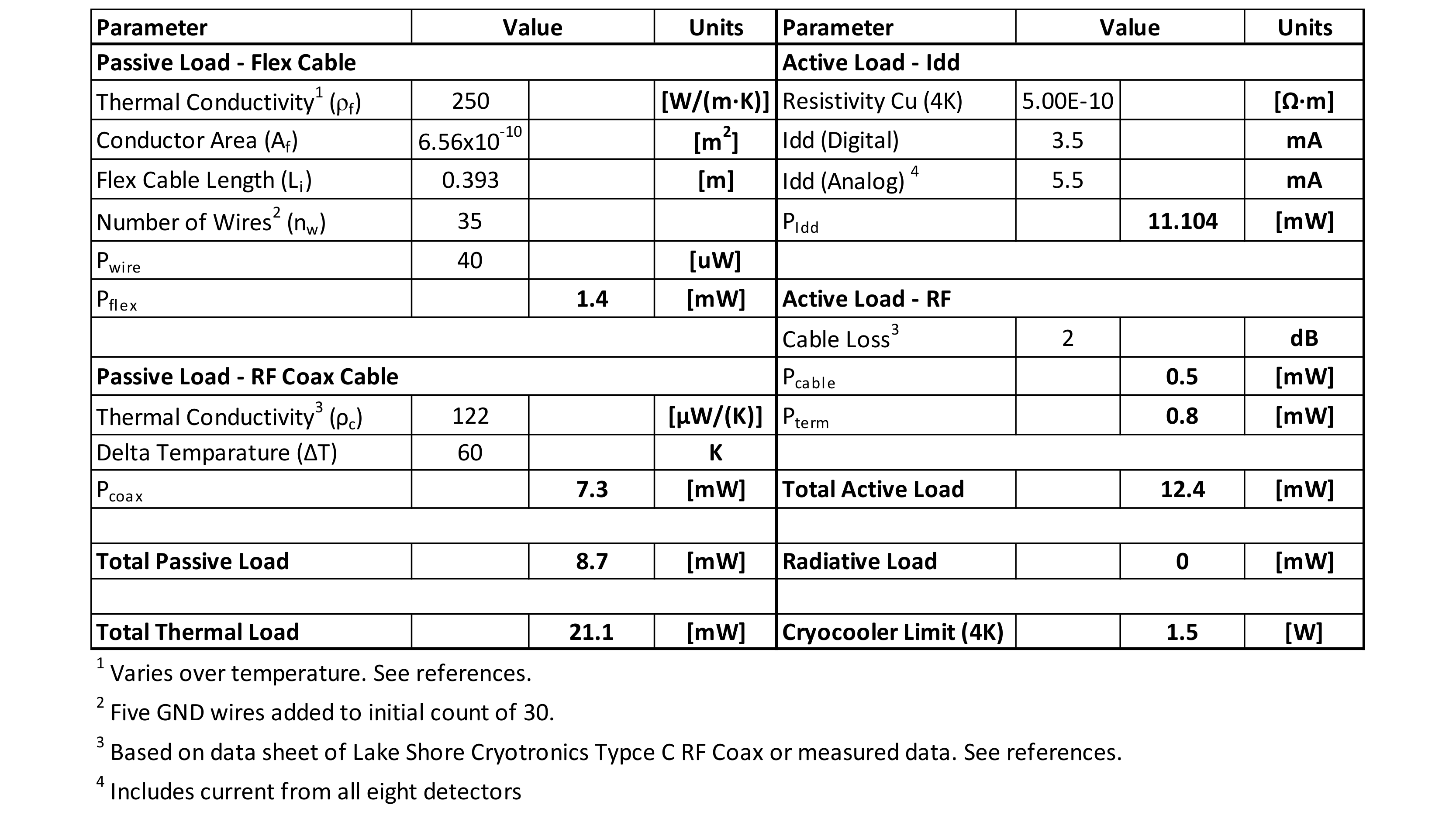}
        \caption{{\bf Thermal load calculations.} The calculations include active and passive loads from cabling connected to DUT including the RF coax carrying the chip clock. Some parameters in the table are taken from ~\cite{Bashir20rfic,nist10}.}. 
        \label{fig:thercalc}
\end{figure}

The heat load calculations, shown in Fig.~\ref{fig:det}, include active and passive loads due to cabling between the DUT PCB and the room-temperature electronics. Radiative loads are neglected in this analysis. The passive loads are calculated for 35 wires (30 from Fig.~\ref{fig:cryosys}(a) plus five ground wires) on the flex cable and one RF coax wire. The thermal conductivity of copper $\rm\ \rho_f(T)$ varies significantly from 3K to 60K~\cite{nist10} and therefore, the analysis calculates the area under the curve over that temperature range to compute the heat transfer due to the flex cable. The equation for the passive thermal load is~\cite{krinner18}: 

\begin{equation}
    P_p= n_w \int_{3K}^{60K} dT \cdot \frac{\rho_f(T) \cdot A_f}{L_f} + \rho_c \cdot \Delta T \
    \label{eq:rhocalc}
\end{equation}

\noindent where $\rm A_f$ and $\rm L_f$ is the cross-section area and length of the conductor on the flex cable respectively, and $\rm n_w$ is the number of wires. The RF coax cable is the Lake Shore Cryotronics Type C cable~\cite{coax} and its thermal conductivity $\rm \rho_c$ is defined in different units than $\rm \rho_f$. The heat transfer from the RF coax is calculated by multiplying by the temperature difference between the cryocooler stages with $\rho_c$. For active load, the square of the average supply current is multiplied by the calculated wire resistance. The heat dissipation from the RF signal is due to the cable loss of 2\,dB~\cite{Bashir20rfic} and the 50\,$\Omega$ termination at the 3K stage. The total thermal load is 21\,mW which is 1.5$\%$ of the 1.5\,W specification. If the 2D qubit array is scaled by a factor of 10 resulting in a 2400 qubit array, the number of detectors will scale to 160. The total thermal load (passive and active) in this case is 150\,mW which is still well below the cryocooler specification. Moreover, on-chip digitization of the detector data will reduce the cabling and consequently the thermal load significantly.

\subsection*{Transfer Function for Noise Analysis at the Injector Output}
\label{sec:inj_tf}

The transfer function $\rm H_{SHPC}$ in Fig.~\ref{fig:cdac}(c) aliases the noise at the switch input and the extent of aliasing is controlled by the duration of pre-charge operation $\rm \tau_{PC}$ and the cycle rate $\rm \tau_{p}$. This equation for $\rm H_{SHPC}$ is determined by modeling the track and hold operation exclusively and assuming the input noise is white over a finite bandwidth~\cite{fischer1982jssc}. During track mode, when the sample switch is closed, the output noise tracks the input. This time duration for this mode, $\rm \tau_{PC}$, is normalized to the switching period:

\begin{equation}
    \tau_{PCn}= \frac{\tau_{PC}}{\tau_p};\\ \tau_{SHn,PC}= 1-\tau_{PCn};\\ f_s= \frac{1}{\tau_{p}}
    \label{eq:tpcn}
\end{equation}

\noindent where $\rm \tau_{SHn,PC}$ is the normalized hold period when the pre-charge switch is open. The input noise from the VDAC is assumed to be white with density $\rm S_{VDAC}$ over a bandwidth $\rm BW_n$ of 500\,MHz used in this analysis. The equation for the transfer function has the same form as Eq.6 in ref.~\cite{fischer1982jssc}:

\begin{equation}
    H_{SHPC}(f) = 2 \tau_{SHn,PC}^{2} \left(\frac{BW_n}{f_s} \right) sinc^{2} \left(\frac{\tau_{SHn,PC}f}{f_s} \right) + (1 - \tau_{SHn,PC})
    \label{eq:hshpc}
\end{equation}

Similarly, the transfer function $\rm H_{SHCDAC}$ models the aliasing of the digital supply noise $\rm S_{VDD}$ to the injector output.

\begin{equation}
    H_{SHCDAC}(f) = 2 \tau_{SHn,clk}^{2} \left(\frac{BW_n}{f_s} \right) sinc^{2} \left(\frac{\tau_{SHn,clk}f}{f_s} \right) + (1 - \tau_{SHn})
    \label{eq:hshcdac}
\end{equation}

\noindent where $\rm \tau_{SHn,clk}$ is the normalized hold period ($\rm \tau_{clk}$=0) for the CDAC path in Fig.~\ref{fig:cdac}(c). 



\subsection*{Detector CDS Transfer Function}
\label{sec:cds_tf}

The CDS transfer function in the detector shown in Fig.~\ref{fig:det}(c) can be analyzed from Eq. 5 in ref.~\cite{pimbley1991tcas}.

\begin{equation}
    H_{CDS}(f) = 4 sinc^2(\pi f T) \cdot \sum_{n=-\infty}^{\infty} S_{in} \left(f - n/T\right)sin^2\left[\pi \lambda \left(f - n/T\right)\right]
    \label{eq:hcds}
\end{equation}

The ``sinc'' term in Eq.~\ref{eq:hcds} is due to the sample and hold operation similar to ref.~\cite{fischer1982jssc}. The summation term of the input spectrum is the aliasing fu
nction. This analysis can be simplified if the input noise spectrum is assumed to be white over the band of interest which is 100\,MHz in the numerical model. With that assumption, the summation term in Eq.~\ref{eq:hcds} can be reduced to an analytical expression based on Eq.~\ref{eq:hshpc}. The replacement term is 2$\frac{BW_n}{fs}$ which essentially is the number of sidebands that are aliased into the bandwidth of interest (see Eq.2 in ref.~\cite{fischer1982jssc}). A simple approach is to structure Eq.~\ref{eq:hcds} as Eq.6 in ref.~\cite{fischer1982jssc} and add the $sin^2\left[\pi \lambda \left(f - n/T \right) \right]$ to account for the difference in two discrete Fourier transforms. Using this approach and the timing parameters shown in Fig.~\ref{fig:det}(c), the CDS transfer function is plotted in Fig.~\ref{fig:det}(d).

\bibliography{refs}

\begin{thebibliography}{10}
\urlstyle{rm}
\expandafter\ifx\csname url\endcsname\relax
  \def\url#1{\texttt{#1}}\fi
\expandafter\ifx\csname urlprefix\endcsname\relax\def\urlprefix{URL }\fi
\expandafter\ifx\csname doiprefix\endcsname\relax\def\doiprefix{DOI: }\fi
\providecommand{\bibinfo}[2]{#2}
\providecommand{\eprint}[2][]{\url{#2}}

\bibitem{knight_2017}
\bibinfo{author}{Knight, W.}
\newblock \bibinfo{title}{\emph{IBM Raises the Bar With a 50-Qubit Quantum
  Computer}}.
\newblock
  \bibinfo{howpublished}{\url{https://www.technologyreview.com/s/609451/
  ibm-raises-the-bar-with-a-50-qubit-quantum-computer}} (\bibinfo{year}{2017}).

\bibitem{JKelley_2017}
\bibinfo{author}{Kelley, J.}
\newblock \bibinfo{title}{\emph{A Preview of Bristlecone, Google's New Quantum
  Processor}}.
\newblock
  \bibinfo{howpublished}{\url{https://ai.googleblog.com/2018/03/apreview-
  of-bristlecone-googles-new.html}} (\bibinfo{year}{2017}).

\bibitem{bogdan2021}
\bibinfo{author}{Staszewski, B.}, \bibinfo{author}{Bashir, I.},
  \bibinfo{author}{Blokhina, E.} \& \bibinfo{author}{Leipold, D.}
\newblock \bibinfo{journal}{\bibinfo{title}{{Cryo-CMOS for Quantum System
  On-Chip Integration}}}.
\newblock {\emph{\JournalTitle{Solid-State Circuits Magazine}}}
  \textbf{\bibinfo{volume}{13}}, \bibinfo{pages}{46--53}
  (\bibinfo{year}{2021}).

\bibitem{krinner18}
\bibinfo{author}{Krinner, S.} \emph{et~al.}
\newblock \bibinfo{journal}{\bibinfo{title}{{Engineering cryogenic setups for
  100-qubit scale superconducting circuit systems}}}.
\newblock {\emph{\JournalTitle{Arxiv: 1806.07862v1}}}  (\bibinfo{year}{2018}).

\bibitem{barends18}
\bibinfo{author}{Barends, R.} \emph{et~al.}
\newblock \bibinfo{journal}{\bibinfo{title}{{Superconducting quantum circuits
  at the surface code threshold for fault tolerance}}}.
\newblock {\emph{\JournalTitle{Nature communications}}}
  \textbf{\bibinfo{volume}{9}}, \bibinfo{pages}{1--6} (\bibinfo{year}{2018}).

\bibitem{bardin_2019b}
\bibinfo{author}{Bardin, J.~C.} \emph{et~al.}
\newblock \bibinfo{journal}{\bibinfo{title}{{Design and Characterization of a
  20-nm Bulk-CMOS Cryogenic Quantum Controller}}}.
\newblock {\emph{\JournalTitle{Solid-State Circuits}}}
  \textbf{\bibinfo{volume}{54}}, \bibinfo{pages}{3043--3060}
  (\bibinfo{year}{2019}).

\bibitem{Brink_2018}
\bibinfo{author}{Brink, M.} \emph{et~al.}
\newblock \bibinfo{title}{Device challenges for near term superconducting
  quantum processors: frequency collisions}.
\newblock In \emph{\bibinfo{booktitle}{IEDM}}, \bibinfo{pages}{6.1.1--3}
  (\bibinfo{year}{2018}).

\bibitem{nikandish21}
\bibinfo{author}{Nikandish, R.}, \bibinfo{author}{Blokhina, E.},
  \bibinfo{author}{Leipold, D.} \& \bibinfo{author}{Staszewski, R.~B.}
\newblock \bibinfo{journal}{\bibinfo{title}{Semiconductor quantum computing:
  Toward a cmos quantum computer on chip.}}
\newblock {\emph{\JournalTitle{IEEE Nanotechnology Magazine}}}
  \bibinfo{pages}{2--14}, \doiprefix\url{10.1109/MNANO.2021.3113216}
  (\bibinfo{year}{2021}).

\bibitem{Maurand_2016}
\bibinfo{author}{Maurand, R.} \emph{et~al.}
\newblock \bibinfo{journal}{\bibinfo{title}{A {CMOS} silicon spin qubit}}.
\newblock {\emph{\JournalTitle{Nature communications}}}
  \textbf{\bibinfo{volume}{7}}, \bibinfo{pages}{13575} (\bibinfo{year}{2016}).

\bibitem{crippa_nature19}
\bibinfo{author}{Crippa, A.}, \bibinfo{author}{Ezzouchm, R.},
  \bibinfo{author}{Aprá, A.} \emph{et~al.}
\newblock \bibinfo{journal}{\bibinfo{title}{{Gate-reflectometry dispersive
  readout and coherent control of a spin qubit in silicon}}}.
\newblock {\emph{\JournalTitle{Nature communications}}}
  \textbf{\bibinfo{volume}{10}}, \bibinfo{pages}{1--6} (\bibinfo{year}{2019}).

\bibitem{zwerver21}
\bibinfo{author}{Zwerver, A. M.~J.} \emph{et~al.}
\newblock \bibinfo{journal}{\bibinfo{title}{{Qubits made by advanced
  semiconductor manfacturing}}}.
\newblock {\emph{\JournalTitle{Arxiv: 2101.12650v}}}  (\bibinfo{year}{2021}).

\bibitem{dijk_2020}
\bibinfo{author}{Van~Dijk, J.~P.} \emph{et~al.}
\newblock \bibinfo{journal}{\bibinfo{title}{{A Scalable Cryo-CMOS Controller
  for the Wideband Frequency-Multiplexed Control of Spin Qubits and
  Transmons}}}.
\newblock {\emph{\JournalTitle{Solid-State Circuits}}}
  \textbf{\bibinfo{volume}{55}}, \bibinfo{pages}{2930--2946}
  (\bibinfo{year}{2020}).

\bibitem{Bonen_2018}
\bibinfo{author}{Bonen, S.} \emph{et~al.}
\newblock \bibinfo{journal}{\bibinfo{title}{Cryogenic characterization of 22-nm
  fdsoi cmos technology for quantum computing ics}}.
\newblock {\emph{\JournalTitle{IEEE Electron Device Letters}}}
  \textbf{\bibinfo{volume}{40}}, \bibinfo{pages}{127--130}
  (\bibinfo{year}{2018}).

\bibitem{ruffino21}
\bibinfo{author}{Ruffino, A.}, \bibinfo{author}{Peng, Y.} \emph{et~al.}
\newblock \bibinfo{title}{{A fully-integrated 40-nm 5-6.5 GHz cryo-CMOS
  system-on-chip with I/Q receiver and frequency synthesizer for scalable
  multiplexed readout of quantum dots}}.
\newblock In \emph{\bibinfo{booktitle}{2021 IEEE International Solid-State
  Circuits Conference (ISSCC)}}, \bibinfo{pages}{210--212}
  (\bibinfo{organization}{IEEE}, \bibinfo{year}{2021}).

\bibitem{Voinigescu_2019}
\bibinfo{author}{{Gong}, M.~J.} \emph{et~al.}
\newblock \bibinfo{title}{Design considerations for spin readout amplifiers in
  monolithically integrated semiconductor quantum processors}.
\newblock In \emph{\bibinfo{booktitle}{2019 IEEE Radio Frequency Integrated
  Circuits Symposium (RFIC)}}, \bibinfo{pages}{111--114}
  (\bibinfo{year}{2019}).

\bibitem{pauka21}
\bibinfo{author}{Pauka, S.~J.} \emph{et~al.}
\newblock \bibinfo{journal}{\bibinfo{title}{{A cryogenic CMOS chip for
  generating control signals for multiple qubits}}}.
\newblock {\emph{\JournalTitle{Nature electronics}}}
  \textbf{\bibinfo{volume}{4}}, \bibinfo{pages}{64--70} (\bibinfo{year}{2021}).

\bibitem{Guevel_2020}
\bibinfo{author}{Guevel, L.~L.} \emph{et~al.}
\newblock \bibinfo{title}{{A 110mK 295µW 28nm {FDSOI} {CMOS} quantum
  integrated circuit with a 2.8{GHz} excitation and nA current sensing of an
  on-chip double quantum dot}}.
\newblock In \emph{\bibinfo{booktitle}{{ISSCC, ses. 19.2}}},
  \bibinfo{pages}{{306--308}} (\bibinfo{year}{{2020}}).

\bibitem{bashir21}
\bibinfo{author}{Bashir, I.} \emph{et~al.}
\newblock \bibinfo{title}{Bias generation and calibration of cmos charge qubits
  at 3.5 kelvin in 22-nm fdsoi}.
\newblock In \emph{\bibinfo{booktitle}{ESSCIRC 2021 - IEEE 47th European Solid
  State Circuits Conference (ESSCIRC)}}, \bibinfo{pages}{47--50},
  \doiprefix\url{10.1109/ESSCIRC53450.2021.9567784} (\bibinfo{year}{2021}).

\bibitem{bashir2020single}
\bibinfo{author}{Bashir, I.} \emph{et~al.}
\newblock \bibinfo{journal}{\bibinfo{title}{A single-electron injection device
  for {CMOS} charge qubits implemented in 22-nm {FD-SOI}}}.
\newblock {\emph{\JournalTitle{IEEE Solid-State Circuits Letters}}}
  \textbf{\bibinfo{volume}{3}}, \bibinfo{pages}{206--209}
  (\bibinfo{year}{2020}).

\bibitem{Gorman_2005}
\bibinfo{author}{Gorman, J.}, \bibinfo{author}{Hasko, D.} \&
  \bibinfo{author}{Williams, D.}
\newblock \bibinfo{journal}{\bibinfo{title}{{Charge-qubit operation of an
  isolated double quantum dot}}}.
\newblock {\emph{\JournalTitle{Physical review letters}}}
  \textbf{\bibinfo{volume}{95}}, \bibinfo{pages}{090502}
  (\bibinfo{year}{2005}).

\bibitem{dkim15}
\bibinfo{author}{Dohun, K.}, \bibinfo{author}{Ward, D.~R.},
  \bibinfo{author}{Simmons, C.~B.} \emph{et~al.}
\newblock \bibinfo{journal}{\bibinfo{title}{{Microwave driven coherent
  operations of a semiconductor quantum dot charge qubit}}}.
\newblock {\emph{\JournalTitle{Arxiv: 1407.7607v1}}}  (\bibinfo{year}{2015}).

\bibitem{Blokhina_2019}
\bibinfo{author}{Blokhina, E.} \emph{et~al.}
\newblock \bibinfo{journal}{\bibinfo{title}{{CMOS} position-based charge
  qubits: Theoretical analysis of control and entanglement}}.
\newblock {\emph{\JournalTitle{IEEE Access}}} \textbf{\bibinfo{volume}{8}}
  (\bibinfo{year}{2020}).

\bibitem{camenzind21}
\bibinfo{author}{Camenzind, L.~C.}, \bibinfo{author}{Geyer, S.},
  \bibinfo{author}{Fuhrer, A.}, \bibinfo{author}{Warburton, R.~J.}
  \emph{et~al.}
\newblock \bibinfo{journal}{\bibinfo{title}{{A spin qubit in a fin field-effect
  transistor}}}.
\newblock {\emph{\JournalTitle{Arxiv: 2103.07369v1}}}  (\bibinfo{year}{2021}).

\bibitem{sscl20-esmailiyan}
\bibinfo{author}{{Esmailiyan}, A.} \emph{et~al.}
\newblock \bibinfo{journal}{\bibinfo{title}{{A Fully Integrated DAC for CMOS
  Position-Based Charge Qubits with Single-Electron Detector Loopback
  Testing}}}.
\newblock {\emph{\JournalTitle{IEEE Solid-State Circuits Letters}}}
  \textbf{\bibinfo{volume}{3}}, \bibinfo{pages}{354--357}
  (\bibinfo{year}{2020}).

\bibitem{fischer1982jssc}
\bibinfo{author}{Fischer, J.~H.}
\newblock \bibinfo{journal}{\bibinfo{title}{{Noise Sources and Calculation
  Techniques for Switched Capacitor Filters}}}.
\newblock {\emph{\JournalTitle{IEEE Journal of Solid State Circuits}}}
  \textbf{\bibinfo{volume}{SC-17}}, \bibinfo{pages}{742--752}
  (\bibinfo{year}{1982}).

\bibitem{chi2011jcas}
\bibinfo{author}{Chi, M.~Y.}, \bibinfo{author}{Christoph, M.} \&
  \bibinfo{author}{Cauwenberghs, G.}
\newblock \bibinfo{journal}{\bibinfo{title}{{Ultra-High Input Impedance, Low
  Noise Integrated Amplifier for Noncontact Biopotential Sensing}}}.
\newblock {\emph{\JournalTitle{Emerging and Selected Topics in Circuits and
  Systems}}} \textbf{\bibinfo{volume}{1}}, \bibinfo{pages}{526--535}
  (\bibinfo{year}{2011}).

\bibitem{pimbley1991tcas}
\bibinfo{author}{Pimbley, J.~M.} \& \bibinfo{author}{Michon, G.~J.}
\newblock \bibinfo{journal}{\bibinfo{title}{{The output power spectrum produced
  by Correlated-Double Sampling}}}.
\newblock {\emph{\JournalTitle{IEEE Transactions of Circuits and Systems}}}
  \textbf{\bibinfo{volume}{38}}, \bibinfo{pages}{1086--1090}
  (\bibinfo{year}{1991}).

\bibitem{esscirc_2019}
\bibinfo{author}{Bashir, I.} \emph{et~al.}
\newblock \bibinfo{title}{A mixed-signal control core for a fully integrated
  semiconductor quantum computer system-on-chip}.
\newblock In \emph{\bibinfo{booktitle}{Proc. of IEEE European Solid-State
  Circuits Conf. (ESSCIRC), ses. A2L-C4}}, \bibinfo{pages}{1--4}
  (\bibinfo{organization}{IEEE}, \bibinfo{year}{2019}).

\bibitem{giounanlis2021cmos}
\bibinfo{author}{Giounanlis, P.} \emph{et~al.}
\newblock \bibinfo{journal}{\bibinfo{title}{{CMOS charge qubits and qudits:
  entanglement entropy and mutual information as an optimization method to
  construct CNOT and SWAP gates}}}.
\newblock {\emph{\JournalTitle{Semiconductor Science and Technology}}}
  (\bibinfo{year}{2021}).

\bibitem{Bashir20rfic}
\bibinfo{author}{Bashir, I.} \emph{et~al.}
\newblock \bibinfo{title}{{RF Clock Distribution System for a Scalable Quantum
  Processor in 22-nm FDSOI Operating at 3.8K Cryogenic Temperature}}.
\newblock In \emph{\bibinfo{booktitle}{2020 IEEE Radio Frequency Integrated
  Circuits (RFIC) Symposium. Digest of Papers, Los Angeles, CA, USA, 2020, pp.
  215--218.}} (\bibinfo{organization}{IEEE}, \bibinfo{year}{2020}).

\bibitem{li_z_scirep18}
\bibinfo{author}{Li, Z.}, \bibinfo{author}{Sotto, M.}, \bibinfo{author}{Liu,
  F.} \emph{et~al.}
\newblock \bibinfo{journal}{\bibinfo{title}{{Random telegraph noise from
  resonant tunnelling at low temperatures}}}.
\newblock {\emph{\JournalTitle{Scientific Reports}}}
  \textbf{\bibinfo{volume}{8}}, \bibinfo{pages}{1--9} (\bibinfo{year}{2018}).

\bibitem{li_z_iedm17}
\bibinfo{author}{Li, Z.} \emph{et~al.}
\newblock \bibinfo{title}{{Random-telegraph-noise by resonant tunnelling at low
  temperatures}}.
\newblock In \emph{\bibinfo{booktitle}{IEDM}}, \bibinfo{pages}{172--174}
  (\bibinfo{year}{2017}).

\bibitem{h_miki12}
\bibinfo{author}{Miki, H.} \emph{et~al.}
\newblock \bibinfo{title}{{Statistical measurement of random telegraph noise
  and its impact in scaled-down high-K metal-gate MOSFETs}}.
\newblock In \emph{\bibinfo{booktitle}{IEDM}}, \bibinfo{pages}{19.1.1--19.1.4}
  (\bibinfo{year}{2012}).

\bibitem{nist10}
\bibinfo{author}{NIST}.
\newblock \bibinfo{title}{\emph{Material Properties: OFHC Copper (UNS
  C10100/C10200)}}.
\newblock
  \bibinfo{howpublished}{\url{https://trc.nist.gov/cryogenics/materials/OFHC\%20Copper/OFHC_Copper_rev1.htm}}
  (\bibinfo{year}{2010}).

\bibitem{coax}
\bibinfo{author}{Shore, L.}
\newblock \bibinfo{title}{\emph{Lake Shore Cryotronics Ultra-maniture cryogenic
  Coaxial Type C Cable}}.
\newblock \bibinfo{howpublished}{\url{https://www.lakeshore.com/ products/
  categories/ overview/ temperature-products/ cryogenic-accessories/
  cryogenic-cable}}.

\end{thebibliography}

\section*{Author contributions statement}

All authors reviewed the manuscript.


\end{document}